\documentclass[useAMS,usenatbib]{mn2e}
\usepackage{cleveref}
\usepackage{natbib}
\usepackage{graphicx}
\usepackage{amsmath}
\usepackage{epstopdf}
\usepackage{subcaption}
\usepackage{float}
\usepackage{xcolor}

\newcommand{\msun}{\,\mathrm{M}_\odot}

\newcommand{\kms}{\,\mathrm{km}\,\mathrm{s}^{-1}}

\newcommand{\au}{\,\mbox{AU}}

\newcommand{\Myr}{\,\mbox{Myr}}
\newcommand{\yr}{\,\mbox{yr}}

\def\lesssim{\mathrel{\hbox{\rlap{\hbox{\lower3pt\hbox{$\sim$}}}\hbox{\raise2pt\hbox{$<$}}}}}
\def\gtrsim{\mathrel{\hbox{\rlap{\hbox{\lower3pt\hbox{$\sim$}}}\hbox{\raise2pt\hbox{$>$}}}}}
\def\gtreq{\mathrel{\hbox{\rlap{\hbox{\lower3pt\hbox{$-$}}}\hbox{\raise2pt\hbox{$>$}}}}}

\title[Turbulence and the formation of circumbinary discs]{The role of turbulence during the formation of circumbinary discs}
\author[Rajika L. Kuruwita, Christoph Federrath]{Rajika L. Kuruwita$^{1}$, Christoph Federrath$^{1}$\\ $^{1}$Research School of Astronomy and Astrophysics, Australian National University, Canberra, ACT 2611, Australia}
\begin{document}
\date{}

\pagerange{\pageref{firstpage}--\pageref{lastpage}} \pubyear{2016}

\maketitle

\label{firstpage}

\begin{abstract}
Most stars form in binaries and the evolution of their discs remains poorly understood. To shed light on this subject, we carry out 3D ideal MHD simulations with the AMR code \texttt{FLASH} of binary star formation for separations of $10-20\au$. We run a simulation with no initial turbulence (NT), and two with turbulent Mach numbers of $\mathcal{M}=\sigma_v/c_s=0.1$ and $0.2$ (T1 and T2) for $5000\yr$ after protostar formation. By the end of the simulations the circumbinary discs in NT and T1, if any, have radii of $\lesssim20\au$ with masses $\lesssim0.02\msun$, while T2 hosts a circumbinary disc with radius $\sim$70-80$\au$ and mass $\sim$0.12$\msun$. These circumbinary discs formed from the disruption of circumstellar discs and harden the binary orbit. Our simulated binaries launch large single outflows. We find that outflows of NT carry the most mass, and linear and angular momentum from the system. T2 produces the least efficient outflows concerning mass, momentum and angular momentum ($\sim$61 per cent, $\sim$71 per cent, $\sim$68 per cent of the respective quantities in NT). We conclude that while turbulence helps to build circumbinary discs which leads to the restructuring of magnetic fields for efficient outflow launching, too much turbulence may disrupt the ordered magnetic field structure required for magneto-centrifugal launching of jets. We conclude that the role of turbulence in building large circumbinary discs may explain some observed very old ($>$10$\Myr$) circumbinary discs. The longer lifetime of circumbinary discs may increase the likelihood of planet formation.
%\thanks{E-mail:
%rajika.kuruwita@anu.edu.au; otheremail@otheraddress (ANO)} and
\end{abstract}

\begin{keywords}
Star Formation -- Binary stars
\end{keywords}

\section{Introduction}
\label{sec:introduction}

A significant fraction of stars are born in binary star systems \citep{raghavan_survey_2010, moe_mind_2017} and in the last two decades a number of planets have been discovered orbiting in binary star systems (e.g. $\gamma$ Cephei Ab \citep{neuhauser_direct_2007}, HD 196885 Ab \citep{chauvin_characterization_2007}, Kepler-47b and c \citep{orosz_kepler-47:_2012}, PH-1 \citep{schwamb_planet_2013}, ROXs 42Bb \citep{kraus_three_2014} and OGLE-2007-BLG-349L(AB)c \citep{bennett_first_2016}). Planet formation models have historically only been concerned with formation around single stars. However, given the frequency of binary stars and the discovery of planets in binaries, this picture is insufficient. Therefore, in order to fully understand planet formation, environments around young binary stars must be considered.

\citet{kraus_impact_2016} find that the frequency of planets orbiting in binary star systems drops for separations of $a\sim 40\au$ \citep{kraus_impact_2016}. However, this work is based on current planet statistics and mostly concerned planets in S-Type orbits, where the planet is orbiting one component in a binary. To date, we only know of $\sim$20 circumbinary planet, thus, we are not able to conduct significant studies of population statistics on these types of planets. In order to understand planet statistics around binary stars we can look at the discs from which these planets would form.

Multiplicity may affect the disc lifetime and as a result can also affect the likelihood of planet formation. The presence of a companion can truncate circumstellar discs leading to the disc material being accreted faster, on the order of $\sim$0.3$\Myr$ \citep{williams_protoplanetary_2011}. A shorter circumstellar disc lifetime is implied by binaries of separation $<40\au$ being half as likely to host circumprimary/circumsecondary discs than binaries with separations $40-400\au$ \citep{cieza_primordial_2009, duchene_planet_2010, kraus_role_2012}, and by close binaries having less sub/millimetre flux due to the absence of an inner disc \citep{jensen_connection_1994, jensen_connection_1996, andrews_circumstellar_2005}.

\cite{harris_resolved_2012} and \cite{cox_protoplanetary_2017} also found that circumstellar discs in binaries produced faint millimetre flux suggesting they are relatively small. However, they also observed circumbinary discs produced at least an order of magnitude more millimetre flux densities compared to circumstellar discs around binaries of the same separation. This implies that circumbinary discs are larger and have more material than circumstellar discs in similar binaries to form planets. There also exist circumbinary discs that are considerably older than the typical lifetime of protoplanetary discs of $\sim$3$\Myr$ \citep{haisch_disk_2001,mamajek_initial_2009}. Examples include AK Sco \citep[$18\pm1\Myr$,][]{czekala_disk-based_2015}, HD 98800 B \citep[$10\pm5\Myr$,][]{furlan_hd_2007}, V4046 Sgr \citep[$12-23\Myr$,][]{rapson_combined_2015} and St 34 \citep[also known as HBC 425, $\sim$25$\Myr$,][]{hartmann_accretion_2005}. If circumbinary discs have a significantly longer lifetime than discs around single stars, it would increase the likelihood of forming planets \citep{kuruwita_multiplicity_2018}.

The mechanisms that determine the lifetime of a disc are: 1. accretion of material, 2. jets and outflows, 3. photo-evaporation of the disc, 4. dynamical interactions. Here we investigate how turbulence influences accretion, outflows and the dynamical evolution of discs during binary star formation and evolution. The work in this paper is a continuation of the work of \cite{kuruwita_binary_2017} which found circumstellar discs were disrupted during the evolution of a young binary star system of separation $\sim$45$\au$, creating a hostile environment for planet formation. In this work we add turbulence to the initial conditions, and run turbulent magnetohydrodynamical simulations of binary star formation to study the evolution of the discs in these systems.

In \Cref{sec:method} we describe the simulation code used, how protostar formation is modelled, our simulation set-up and implementation of turbulence. The results are presented and discussed in \Cref{sec:results}, where we analyse the evolution of the binary systems, the outflows produced and the evolution of the discs. \Cref{sec:caveats} discusses the limitations and caveats of this study. Our conclusions are summarised in \Cref{sec:conclusion}.

\section{Method}
\label{sec:method}

\subsection{\texttt{FLASH}}
\label{ssec:flash}

The simulations are carried out with \texttt{FLASH} which is a magnetohydrodynamic (MHD) adaptive mesh refinement (AMR) code \citep{fryxell_flash:_2000, dubey_challenges_2008}. \texttt{FLASH} integrates the ideal MHD equations. We use the HLL3R Riemann solver for ideal MHD \citep{waagan_robust_2011}. The gravitational interactions of the gas are calculated using a Poisson solver \citep{wunsch_tree-based_2018}.

Our simulations use a piecewise-polytropic equation of state, given by:

\begin{equation}
P_\mathrm{th} = K\rho^\Gamma,
\label{eqn:eos}
\end{equation}

\noindent where $K$ is the polytropic constant and $\Gamma$ is the polytropic index. For our simulations $\Gamma$ is defined to be:

\begin{equation}
\Gamma=\begin{cases}
1.0  \text{ for \,\,\,\,\,\,\,\,\,\,\,\, $\rho \leq \rho_1 \equiv 2.50 \times 10^{-16}\,\mathrm{g}\,\mathrm{cm}^{-3}$},\\
1.1  \text{ for $\rho_1 < \rho \leq \rho_2 \equiv 3.84 \times 10^{-13}\,\mathrm{g}\,\mathrm{cm}^{-3}$},\\
1.4  \text{ for $\rho_2 < \rho \leq \rho_3 \equiv 3.84 \times 10^{-8}\,\mathrm{g}\,\mathrm{cm}^{-3}$},\\
1.1  \text{ for $\rho_3 < \rho \leq \rho_4 \equiv 3.84 \times 10^{-3}\,\mathrm{g}\,\mathrm{cm}^{-3}$},\\
5/3 \text{ for \,\,\,\,\,\,\,\,\,\,\,\,$\rho > \rho_4$}.
\end{cases}
\label{eqn:gamma}
\end{equation}

These values were determined by radiation-hydrodynamical simulations of molecular core collapse by \citet{masunaga_radiation_2000}. These values of $\Gamma$ approximate the gas behaviour during the initial isothermal collapse of the molecular core, adiabatic heating of the first core, the H$_2$ dissociation during the second collapse into the second core and the return to adiabatic heating. 

The formation of sink particles indicates the formation of a protostar \citep{federrath_modeling_2010, federrath_implementing_2011, federrath_modeling_2014}. A second-order leapfrog integrator is used to update the particle positions using a velocity and acceleration to calculate a variable time step. To prevent artificial precession of the sink particles a sub-cycling method is implemented \citep{federrath_modeling_2010}. The interactions between sink particles and the gas are computed using $N$-body integration.

For further details about \texttt{FLASH} and sink particles, refer to \citet{kuruwita_binary_2017} and references therein.

\subsection{Simulation setup}
\label{ssec:simulationsetup}

\begin{figure*}
	%\centerline{\includegraphics[width=1.0\linewidth]{Images/Side_on_volume_weighted.pdf}}
	\centerline{\includegraphics[width=1.0\linewidth]{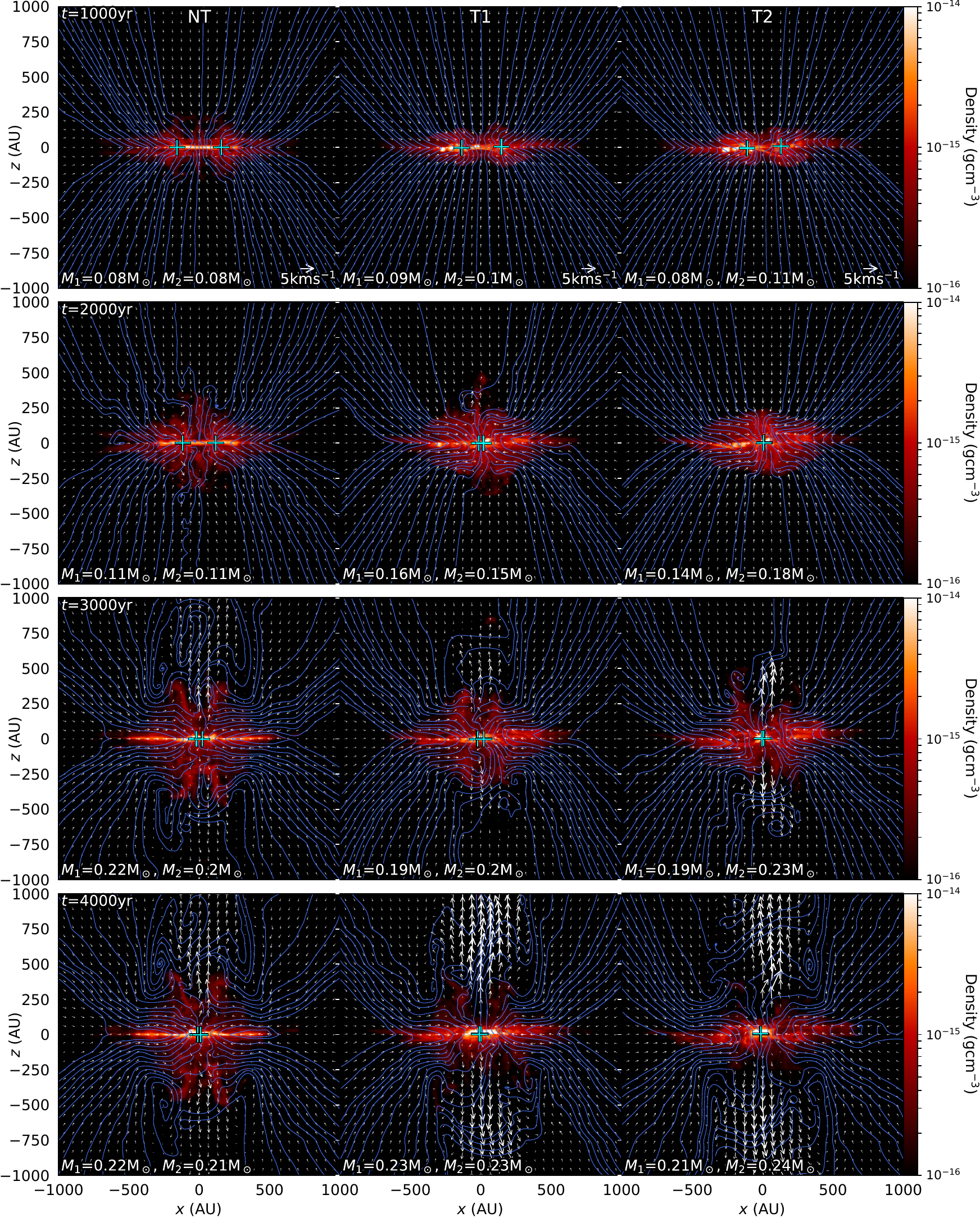}}
	\caption{$300\au$ thick volume-weighted slices orientated along the dense accretion streams such that the slice captures the dense material and two sink particles for NT (left), T1 (middle) and T2 (right). Each row progresses at $1000\yr$ intervals since the first protostar formation. The thin lines show the magnetic field, and the arrows indicate the velocity field. Crosses show the position of the sink particles. The mass accreted by the sink particles in the simulations is indicated on the bottom left of each panel.}
	\label{fig:side_on_slices}
\end{figure*}

\begin{figure*}
	%\centerline{\includegraphics[width=1.0\linewidth]{Images/Top_down.pdf}}
	\centerline{\includegraphics[width=1.0\linewidth]{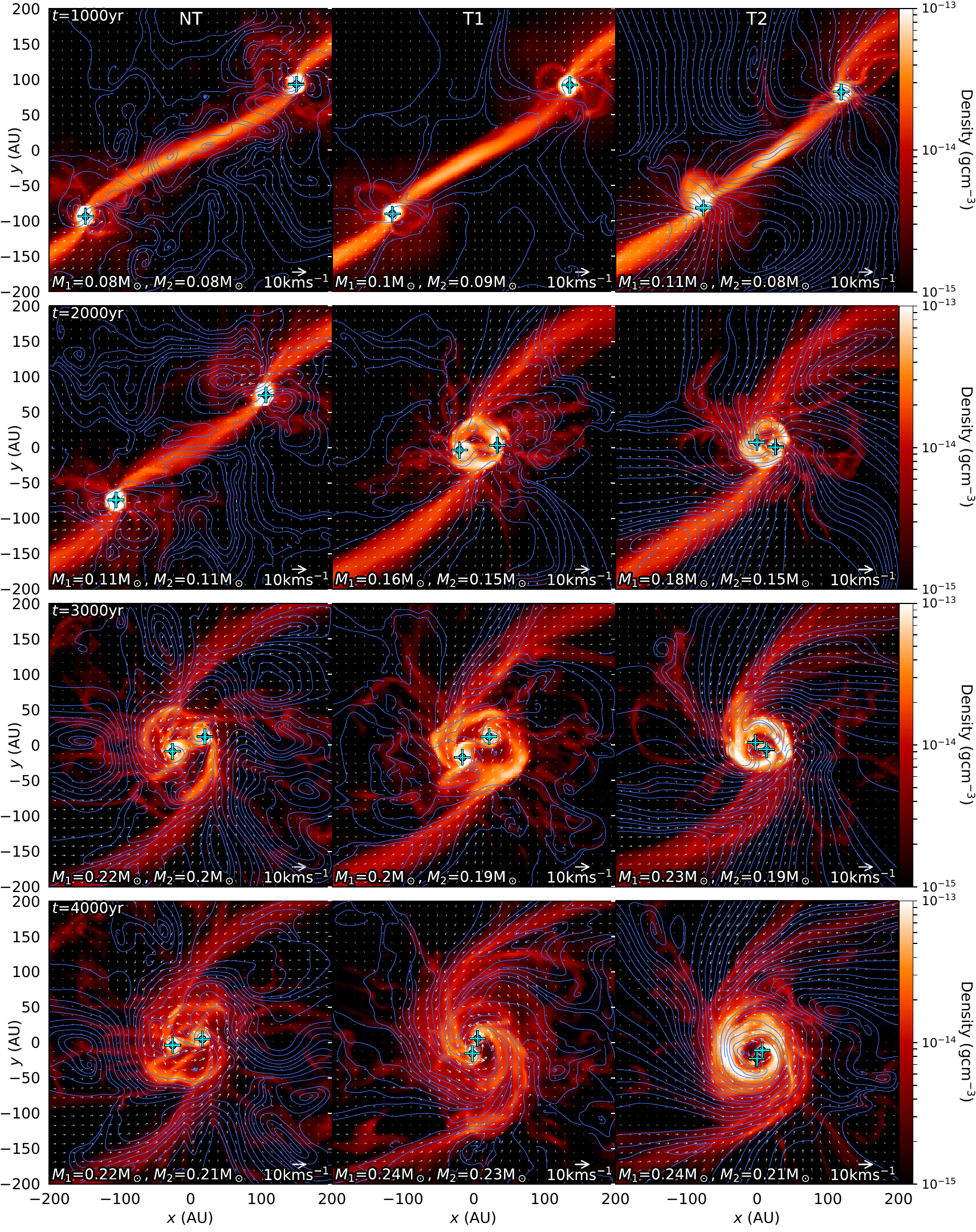}}
	\caption{Same as \Cref{fig:side_on_slices}, but for $100\au$ thick slices along the $z=0$ plane of the simulation box.}
	\label{fig:top_down_slices}
\end{figure*}

The simulations methods are identical to \citet{kuruwita_binary_2017}, except for adding turbulence (see details in \Cref{ssec:addingturbulence}). Here we only summarise the main elements of the method and refer the reader to \citet{kuruwita_binary_2017} for the details. We simulate the formation of a binary star with turbulence and without turbulence for comparison of outflow quantities and evolution of environment around the protostars.

The size of the computational domain is $L_\mathrm{box}=1.2\times10^{17}\,\mathrm{cm}$ ($\sim$8000$\au$) along each side of the 3D computational domain. On the highest level of refinement of the AMR grid the resolution is $\sim$1.95$\au$. At this resolution the accretion radius of the sink particles is $r_\mathrm{sink}\sim$4.9$\au$.

Our simulations begin with a spherical cloud of mass $1\msun$, and radius $\sim$3300$\au$ placed in the centre of the simulation domain. In the non-turbulent case the cloud is given solid body rotation with angular momentum of $1.85\times10^{51}\,\mathrm{g}\,\mathrm{cm}^2\,\mathrm{s}^{-1}$. With this angular momentum, the product of the angular frequency and the free-fall time of the cloud is $\Omega\times\,t_\mathrm{ff}=0.2$ (see \cite{banerjee_outflows_2006} and \cite{machida_high-_2008}). A magnetic field of $10^{-6}\,\mathrm{G}$ is also threaded through the cloud in the \emph{z}-direction. This gives a mass-to-flux ratio of $(M/\Phi)/(M/\Phi)_{\mathrm{crit}}=5.2$ where the critical mass-to-flux ratio is $487\,\mathrm{g}\,\mathrm{cm}^{-2}\,\mathrm{G}^{-1}$ as defined in \citet{mouschovias_note_1976}. 

The cloud is initially given a uniform density of $\rho_0=3.82\times10^{-18}\,\mathrm{g}\,\mathrm{cm}^{-3}$ and then a density perturbation is imposed on the cloud. This is to seed the formation of a binary star system. Previous works of multiple star formation from a single molecular core suggest that cloud fragmentation is a more likely pathway of multiple star formation rather than disc fragmentation \citep{offner_formation_2010}. This is due to radiation feedback increasing the Jeans length within discs, suppressing fragmentation. Our initial conditions are designed such that our binary star systems form from core fragmentation in agreement with previous results. The density perturbation in our simulations is described by:

\begin{equation}
\rho = \rho_0 (1 + \alpha_\mathrm{p}\mathrm{cos}\phi),
\label{eqn:densityperturbation}
\end{equation}

\noindent where $\phi$ is the angle about the $z$-axis and $\alpha_\mathrm{p}$ is the amplitude of the perturbation. For our simulations $\alpha_\mathrm{p}=0.50$. This perturbation is a standard method of seeding binary star formation within simulations of molecular cores \citep{boss_fragmentation_1979, bate_resolution_1997}.

In order to prevent the cloud from expanding rapidly, the spherical cloud is in pressure equilibrium with the surrounding material. This is set by giving the surrounding material a gas density of $\rho_0/100$ with an internal energy such that the cloud and surrounding material is in pressure equilibrium. Inflow/outflow boundary conditions are used at the edge of our computational domain.

\subsubsection{Adding turbulence}
\label{ssec:addingturbulence}

\begin{table}
	\centering
	\begin{tabular}{lccc}
		\hline
		Simulation name & $\sigma_{v}$ ($\mathrm{km}\,\mathrm{s}^{-1}$) & $\mathcal{M}$ & $E_{turb}$ (erg)\\
		\hline
		NT & $0.0$ & $0.0$ & $0.0$\\ 
		T1 & $0.02$ & $0.1$ & $4.0\times10^{39}$\\
		T2 & $0.04$ & $0.2$ & $1.6\times10^{40}$\\ 
		\hline
	\end{tabular}
	\caption{Summarises the simulations. The left column gives the simulation name. The right column gives the amount of initial turbulence ($\sigma_{v}$) as described by Equation \ref{eqn:turbulentenergy}.}
	\label{tab:simulation_summary}
\end{table}

Stars form in turbulent cores inside molecular clouds \citep{ferriere_interstellar_2001, mac_low_control_2004, padoan_infall-driven_2014}. In order to simulate this effect, we add turbulence to the initial conditions of our cores. The purpose of this work is to investigate the role of turbulence on disc evolution during binary star formation. Therefore to isolate the effect of turbulence all simulations in this work have the same initial conditions bar the level of turbulence introduced to the initial velocity field of the cloud.

For our simulations we derive a ball-park value for the velocity dispersion by setting the turbulent energy to a fraction of the rotational energy. The turbulent energy is given by the following:

\begin{equation}
E_{turb} = \frac{1}{2}M\sigma_v^2,
\label{eqn:turbulentenergy}
\end{equation}

where $M$ is the initial mass of our cloud i.e. $1\msun$ and $\sigma_v$ is the velocity dispersion. The initial rotational energy of the cloud in our simulations is $E_{rot}=3.4\times10^{40}\,\mathrm{erg}$. Here we study $\sigma_{v}=0,0.02$ and $0.04\kms$ or Mach 0.0, 0.1, and 0.2 which corresponds to $E_{turb}/E_{rot}=0.0, \sim$0.12 and $\sim$0.46. This is consistent with mild, subsonic levels of turbulence expected to be present in star-forming cores and discs. Hereafter the simulations with turbulence of $\mathcal{M}=0.0,0.1$ and $0.2$ are referred to as NT, T1 and T2 respectively. The simulations are summarised in \Cref{tab:simulation_summary}. 

The mixture of solenoidal and compressible modes is set such that the velocity field only contains solenoidal modes initially, i.e., the velocity field is divergence-free \citep{federrath_comparing_2010}. The spectrum of initial turbulent velocities follows a \citet{kolmogorov_local_1941} spectrum with a spectral slope of $k^{-5/3}$ (this is the expected power-law slope for the mildly compressible regime of turbulence relevant here; see \citet{federrath_universality_2013}), populated in the wavenumber range $k=[2,20]$, where $k$ is in units of $2\pi/L_\mathrm{box}$.

%For our work we decided to run simulations with turbulence of $\mathcal{M} = 0.1$ and $0.2$. Greater levels of turbulence resulted in further fragmentation, which is not suitable for our study of binary star formation.

The random seed for the turbulence is the same for all simulations. The turbulence is not driven in our simulations, it is only applied to the initial velocity field of the gas during the set-up. 

\section{Results and discussion}
\label{sec:results}

Our simulations were run and we followed the system evolution. Side-on and top-down slices of the resulting simulations are shown in \Cref{fig:side_on_slices} and \Cref{fig:top_down_slices}. \Cref{fig:side_on_slices} shows side-on gas density slices of thickness $300\au$ at $1000\yr$ intervals since the formation of the first sink particle. \Cref{fig:top_down_slices} shows top-down gas density slices of thickness $100\au$ at $1000\yr$ intervals since the formation of the first sink particle. The orientation of the side-on slices in \Cref{fig:side_on_slices} is aligned along the dense accretion streams we see in \Cref{fig:top_down_slices}. 

%First we look at the evolution of the formed protostars in \Cref{ssec:evolution}. Secondly we look at the outflows from the systems in \Cref{ssec:outflows}. Finally we look at the dynamical evolution of the discs in these systems in \Cref{ssec:discstructure}.

\subsection{Time evolution of the binary star system}
\label{ssec:evolution}

\begin{figure}
	%\centerline{\includegraphics[width=1.0\linewidth]{Images/binary_system_time_evolution.pdf}}
	\centerline{\includegraphics[width=1.0\linewidth]{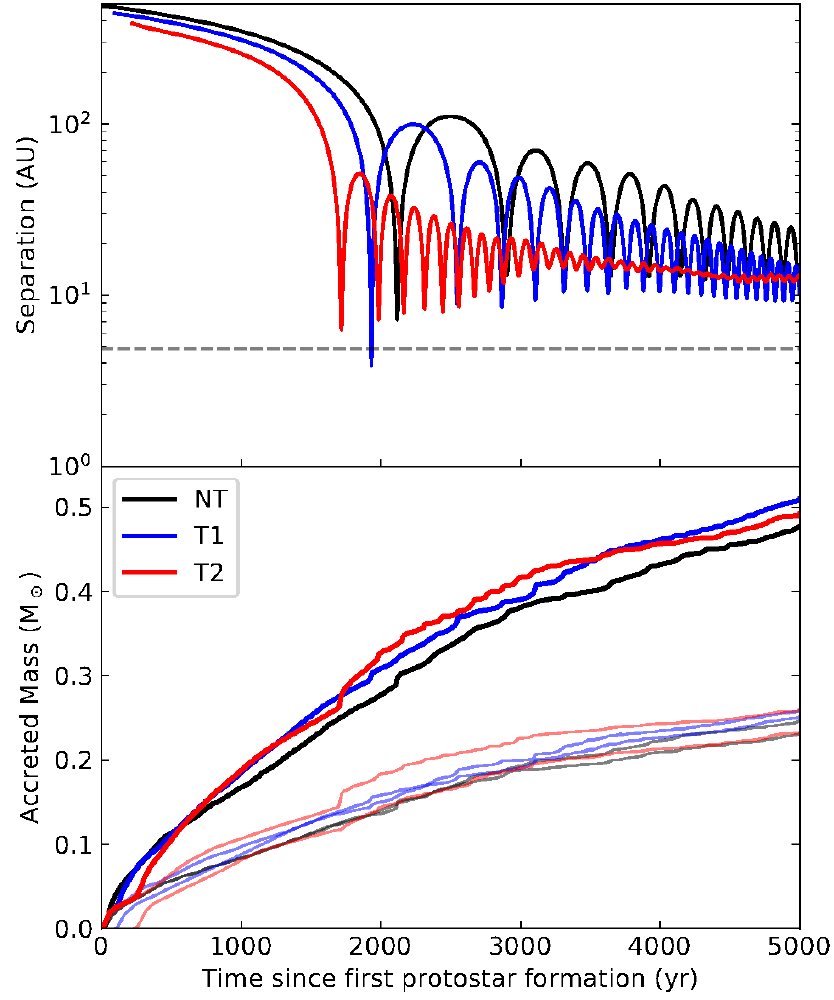}}
	%\caption{Shows the evolution of the binary separation (\emph{Top}), the total accretion rate (\emph{Middle}) and accreted mass (\emph{Bottom}) since protostar formation of the first sink particle for the NT, T1 and T2 cases in black, blue and red respective. In the plot of the accreted mass, the transparent thin lines trace the mass of the individual components and the thick opaque lines traces the total accreted mass.}
	\caption{Shows the evolution of the binary separation (top), and accreted mass (bottom) since protostar formation of the first sink particle for the NT, T1 and T2 cases in black, blue and red respective. In the plot of the accreted mass, the transparent thin lines trace the mass of the individual components and the thick opaque lines traces the total accreted mass.}
	\label{fig:SystemEvolution}
\end{figure}

As the simulations progress, the spherical cloud collapses and sink particles are created in collapsing regions (c.f. \Cref{ssec:flash}). Sink particles form at separations between $400-500\au$ and fall towards the centre of the initial dense core as shown in \Cref{fig:top_down_slices} and the top plot of \Cref{fig:SystemEvolution}. The initial separation and delay between the formation of the first and second sink particle is dependant upon the strength of the turbulence of the initial velocity field (c.f. \Cref{ssec:addingturbulence}). In the non-turbulent NT simulation the sink particles form sooner and at the same time because the density perturbation is symmetric. In T1 and T2 the turbulence washes out some of the initial m=2 density perturbation (refer to $\alpha_\mathrm{p}$ in Equation \ref{eqn:densityperturbation}) and the sink particles form at later absolute simulation times as a result. The turbulence in T1 and T2 also introduces asymmetries to the initial density perturbation creating a delay between the formation of the sink particles, producing binary systems with unequal mass components. The stronger the turbulence the greater the delay between formation of the sink particles. Our binary systems are evolved for $5000\yr$ after formation of the first sink particle. This ensured that the binary was able to complete many orbits to form an established binary system of semi-major axis between $\sim$10-20$\au$. The dashed line in \Cref{fig:SystemEvolution} (top panel) indicates the accretion radius of our sink particles (which is $r_{\mathrm{sink}}=4.9\au$) to demonstrate that the separation of the binaries is not limited by numerical resolution. The binary systems begin to establish their orbits approximately $\sim$2000$\yr$ after the formation of the first sink particle in all cases. We see that T2 has fully circularised approximately $\sim$4000$\yr$ after the first sink particle formation, while in NT and T1, circularisation is still ongoing after 5000$\yr$. This orbital shrinkage after protostar formation has also been seen in other MHD simulations of binary star formation and is caused by magnetic braking of the infalling gas \citep{zhao_orbital_2013}.

As the binary evolves the sink particles accrete mass. The bottom panel of \Cref{fig:SystemEvolution} shows the mass accreted by the sink particles in our three simulations. The thick solid lines show the total accreted mass and the mass of the individual sink particles is indicated by the thin transparent lines. The mass ratio ($M_{secondary}/M_{primary}$) is lowest for the most turbulent simulation. This is primarily due to the time delay between the formation of each component. We see that the turbulent cases have a greater star formation efficiency (fraction of accreted mass) than NT at all times, but the amount of accreted mass does not very significantly. $5000\yr$ after sink particle formation, T1 and T2 have respectively accreted $\sim$8$\%$ and $3\%$ more mass than NT. We cannot predict the final mass of the stars, because it is impossible to run these simulations until the stars stop accreting mass, due to the limited amount of compute time currently available.

Initially the accretion rate is very high, but steadily falls over the course of our simulations. All simulations show episodes of increased accretion rate around the first few periastron passages. This supports the binary trigger hypothesis for accretion events leading to FU Ori type outbursts \citep{green_testing_2016, tofflemire_pulsed_2017}. %However after a few orbits, it is difficult to correlate times of high accretion with the timing of periastron passage. This is likely tied to the environment around the sink particles at these later times. 

The accretion rate of NT decreases significantly after the third periastron passage at $\sim$3000$\yr$. We see from \Cref{fig:top_down_slices} that the circumstellar discs of NT are heavily disrupted and smaller at times after $3000\yr$. The circumstellar discs help funnel gas onto the sink particles. Because these discs are disrupted, the accretion rate onto the sink particles is reduced.

In contrast we also see the average accretion rate of T2 decreases at approximately the time when the binary system has circularised between $3000-4000\yr$. We see from \Cref{fig:top_down_slices} that for T2, a dense circumbinary disc is building up and the magnetic field is coiling up in this disc, while a cavity appears to open near the binary stars at times after $3000\yr$. The establishment of the circumbinary disc helps circularise the binary system by removing angular momentum from the binary orbit and transporting that into the disc. The reduced eccentricity of the orbit also means the cross-section in which mass must exist to be accreted is reduced. The low density of material near the sink particles at later times also affects the accretion rate.

We now turn to analysing the properties of the outflows/jets launched from these binary-star systems.

\subsection{Morphology of outflows and launching mechanism}
\label{ssec:outflows}

\Cref{fig:side_on_slices} qualitatively shows the outflow morphology in the three cases. In NT the outflows from the sinks are launched at the same time and remain symmetric over the course of the simulation. T1 produces asymmetric outflows because the turbulence induces asymmetric density perturbations. There is even greater delay between the formation of the first and second sink particle in T2, however in this case we do not see jets from the individual sink particles. This may be due to the stronger turbulence creating an unordered magnetic field reducing the magnetic pressure gradient required for efficient launching.

Individual jets from young proto-binaries have been observed, such as L1551 IRS5 \citep{wu_multiple_2009}. The morphology of the outflows from this system is similar to those produced by the NT and T1 cases. The duration of the individual jets from our simulations depends on the separation and coalescence time of the binary star system. The duration of the individual jets in our simulations is of the order of $1000-2000\,\mathrm{yr}$. The L1551 IRS5 system has a separation of $\sim$42$\,\mathrm{AU}$ \citep{bieging_multifrequency_1985} which is $2-3$ times larger than the binaries produced in our simulations. This system could be in-spiralling and the duration of these individual outflows is also limited by the time taken for the circumstellar discs to be disrupted by dynamical interactions. The duration of these individual outflows can be assumed to depend on the separation of the binary system and the detection of these double outflows can help trace recent binary star formation. However, binaries formed in more turbulent environments may be harder to detect in the early stages of formation as individual jets are not produced as shown by T2. This would mean that binary star formation in very turbulent environments can be obscured from observations.

In all cases, once the binary has been established after many orbits the system produces a single outflow. The speed of the outflows increases at later times, with the more turbulent cases producing greater bulk outflow speeds.

\begin{figure*}
	%\centerline{\includegraphics[width=1.0\linewidth]{Images/binary_system_time_evolution.pdf}}
	\centerline{\includegraphics[width=1.0\linewidth]{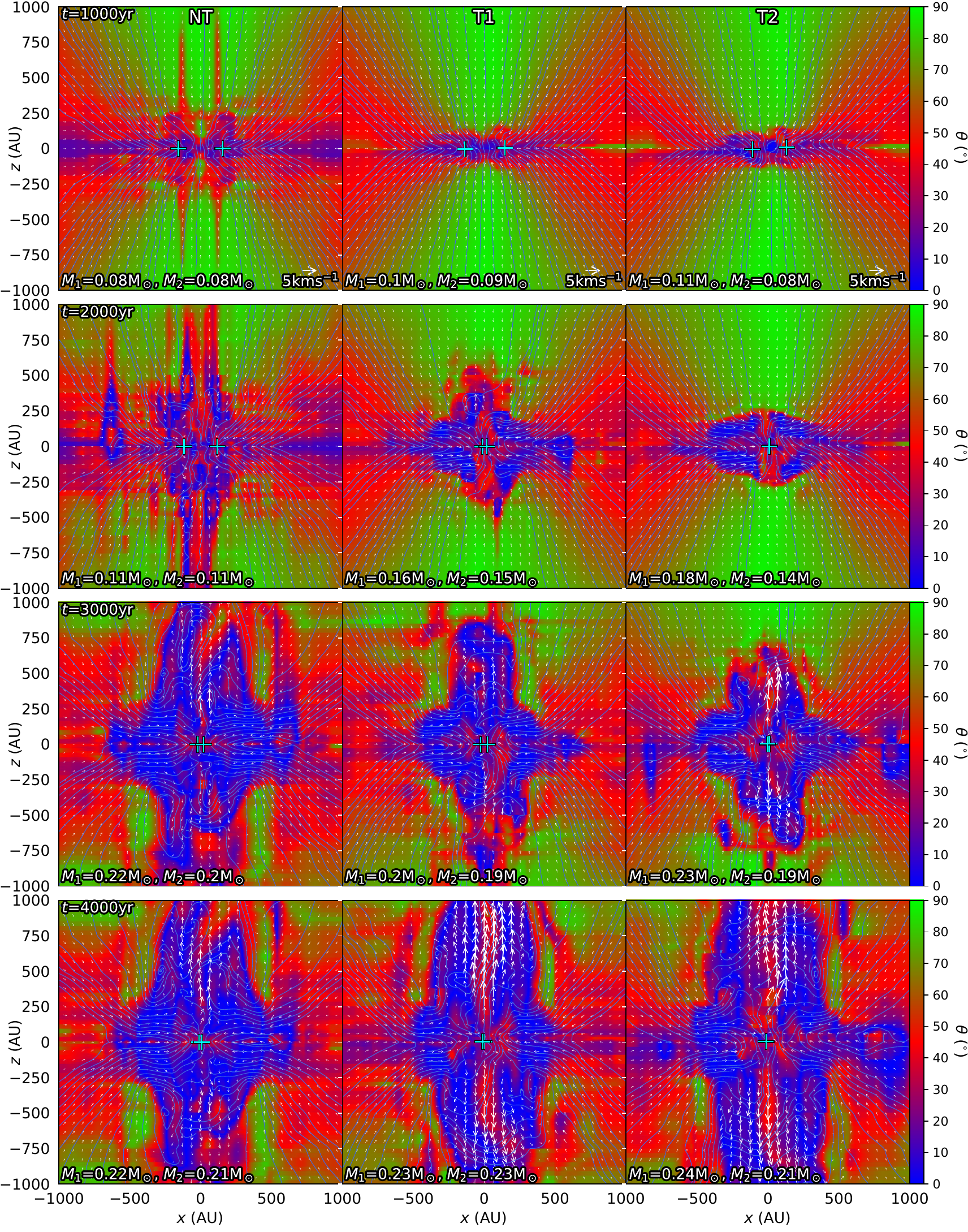}}
	%\caption{Shows the evolution of the binary separation (\emph{Top}), the total accretion rate (\emph{Middle}) and accreted mass (\emph{Bottom}) since protostar formation of the first sink particle for the NT, T1 and T2 cases in black, blue and red respective. In the plot of the accreted mass, the transparent thin lines trace the mass of the individual components and the thick opaque lines traces the total accreted mass.}
	\caption{Same as \Cref{fig:side_on_slices}, but shows volume-weighted slices of the magnetic field angle to the $xy$-plane.}
	\label{fig:angle}
\end{figure*}

\begin{figure*}
	%\centerline{\includegraphics[width=1.0\linewidth]{Images/binary_system_time_evolution.pdf}}
	\centerline{\includegraphics[width=1.0\linewidth]{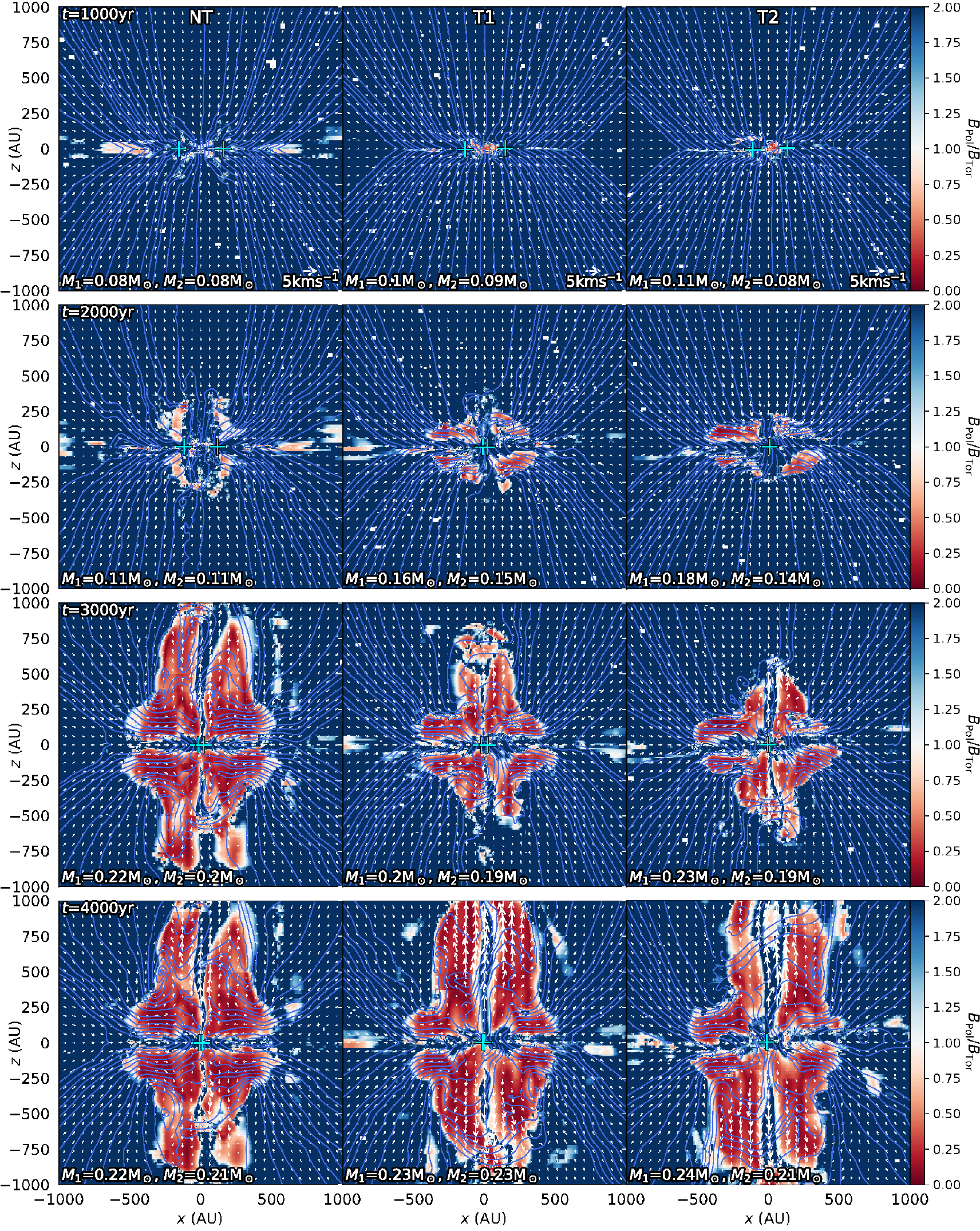}}
	%\caption{Shows the evolution of the binary separation (\emph{Top}), the total accretion rate (\emph{Middle}) and accreted mass (\emph{Bottom}) since protostar formation of the first sink particle for the NT, T1 and T2 cases in black, blue and red respective. In the plot of the accreted mass, the transparent thin lines trace the mass of the individual components and the thick opaque lines traces the total accreted mass.}
	\caption{Same as \Cref{fig:side_on_slices}, but shows volume-weighted slices of the ratio of the poloidal component ($B_\mathrm{pol}$) to the toroidal component ($B_\mathrm{tor}$) of the magnetic field.}
	\label{fig:pol_to_tor_ratio}
\end{figure*}

Based on the magnetic field structure in the simulations we attribute the driving force of the outflows to the magneto-centrifugal mechanism. \cite{blandford_hydromagnetic_1982} calculated that gas can be launched centrifugally along magnetic field lines if the poloidal component of the magnetic field ($B_\mathrm{pol}$) subtends an angle of less than $60^\circ$ from the disc surface. Once an outflow is launched, the toroidal component of the magnetic field ($B_\mathrm{tor}$) becomes important and collimates the outflow. To test this we calculate both the angle of the magnetic field to the $xy$-plane ($\theta$) and the ratio of the poloidal to toroidal magnetic field components. Slices of the magnetic field angle and the poloidal to toroidal ratio are shown in Figures \ref{fig:angle} and \ref{fig:pol_to_tor_ratio}.

From \Cref{fig:angle} we see at early times the magnetic field angle is shallow in the midplane ($<30^\circ$) and is steep away from the disc ($>60^\circ$). As the simulations progress the transition region where the magnetic field angle transitions from $>60^\circ$ to $<60^\circ$ follows the outflow fronts seen in \Cref{fig:side_on_slices}. From \Cref{fig:pol_to_tor_ratio} we also see at early time the poloidal component dominates the magnetic field structure. However, at later times after the outflows are launched, we find that the toroidal component grows and begins to dominate the magnetic field around the outflows. It is this toroidal field that collimates the outflows. We also see that the poloidal field is the primary component in the disc midplane. These results indicate that these outflows are launched via the magneto-centrifugal mechanism and are in line with other simulations that also show this \citep{banerjee_outflows_2006}.

\subsection{Measurement of the outflows}
\label{ssec:outflow_measurement}

Mass, momentum, and angular momentum can be carried away from protostellar systems via outflows. Here we measure these outflow quantities in our three scenarios to determine how turbulence may affect outflows from young binary star systems.

The scale height of the discs is approximately $z_\mathrm{H}\approx25\au$ in the non-turbulent case. Analysis of the outflows from the systems is carried out by measuring the outflowing mass within two measuring cylinders placed above and below the $z=0$ plane. The face of the cylinder nearest the discs is placed $|z|\geq 4z_\mathrm{H} = 100\au$ from the $z=0$ plane. This height is selected to capture the outflow material, while avoiding measuring turbulent material in the warped discs in T1 and T2. The radius  and height of the measuring cylinders is $500$ and $3300\au$ respectively to capture and track outflowing material. Within the measuring regions outflow mass is defined as any mass in cells with $v_z>0$ for $z>0$ and $v_z<0$ for $z<0$. From the outflow mass, the angular momentum and linear momentum of the outflows is calculated, as well as the maximum outflow speed. The linear momentum is calculated from the magnitude of the velocity from the centre of mass and the outflow mass. The angular momentum is calculated about the centre of mass of the systems. 

\begin{figure}
	%\centerline{\includegraphics[width=1.0\linewidth, trim={4mm, 0.0mm, 10.0mm, 0.0mm}]{Images/outflow_quantities.pdf}}
	\centerline{\includegraphics[width=1.0\linewidth]{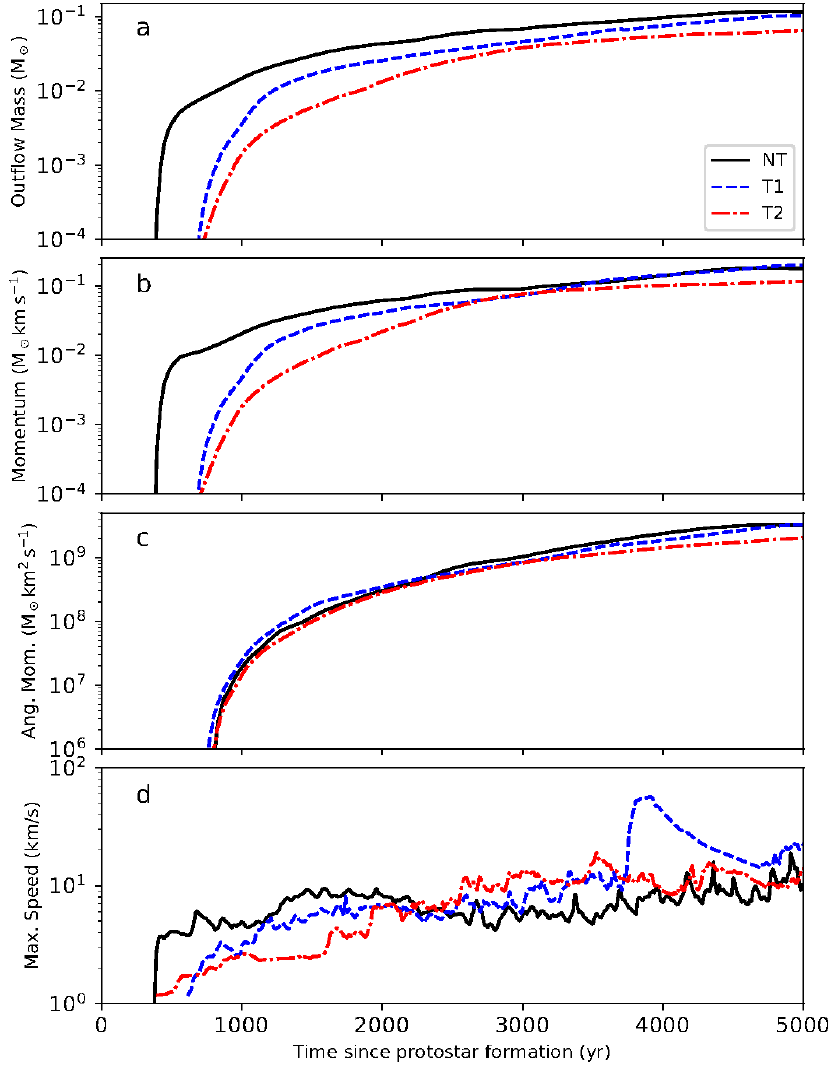}}
	\caption{Time evolution of the outflow quantities measured from cylinders of radius $500\au$ and height $500\au$ placed at $100\au $ above and below the disc $z=0$ plane. Panel a: outflowing mass, defined as mass within cells with $v_z$ away from the disc. Panel b: linear momentum of the outflowing gas. Panel c: angular momentum of the outflowing gas calculated around the centre of mass of the system. Panel d: the speed of the cell with the maximum speed within the measuring volume.}
	\label{fig:outflow_quantities}
\end{figure}

The outflow measurements of the three cases is shown in \Cref{fig:outflow_quantities}. Most of the measured quantities reach a plateau as the quantity flowing into the measuring volumes is equal to the amount of the quantity leaving the volumes. From \Cref{fig:outflow_quantities} we see that the outflows of NT reach the measuring volume the fastest. This is also confirmed in the bottom plot of \Cref{fig:outflow_quantities} where NT has the highest initial outflow velocity. While these curves shown in \Cref{fig:outflow_quantities} illustrate the rate at which these quantities are carried in the outflows, to compare the overall outflow efficiency of these quantities we calculate the time averaged values of the quantity using:

\begin{equation}
<q>=\frac{\int_{0}^{T}q(t)dt}{\int_{0}^{T}dt},
\label{eqn:time_averaged}
\end{equation}

\noindent where $q$ is mass, momentum, and angular momentum, and $T=5000\yr$.

Panel a of \Cref{fig:outflow_quantities} shows the mass carried in the outflows. We see that NT is the most efficient at transporting mass via outflows and T2 is the least efficient. T1 falls between the other two cases with respect to the outflow mass. The time averaged values of the outflow mass for T1 and T2 are $\sim$90$\%$ and $\sim$61$\%$ of NT, respectively.

Initially the outflow of NT carries the greatest linear momentum. This is mostly due to having the most massive outflows and faster outflow velocities. However at later times T1 converges towards NT. This may be due to increased outflow velocities. The outflows of T2 carry the least amount of linear momentum over the course of the entire simulation due to having the least massive outflows and have lower outflow velocities. The time averaged values of the linear momentum carried in the outflows for T1 and T2 are $\sim$111$\%$ and $\sim$71$\%$ of NT, respectively.

All the simulations show similar outflow behaviours for angular momentum. The time averaged values of the angular momentum  carried in the outflows for T1 and T2 are $\sim$102$\%$ and $\sim$68$\%$ of NT, respectively.

The maximum outflow varies between the three cases. In NT the maximum outflow speed remains relatively steady over the course of the entire simulation, sitting at the level of $3-10\kms$. There are bursts in the outflow velocity for NT which happen around the time of periastron passage as seen in \Cref{fig:SystemEvolution}. The maximum outflow velocity of T1 grows to a steady state by $\sim$1000$\yr$, with speeds in the same range as the NT case. At $\sim$3800$\yr$ there is a sharp rise in the outflow velocity. From \Cref{fig:top_down_slices}, between times $3000$ and $4000\yr$, we see that the magnetic field structure in T1 has become mostly organised, coiling around the binary. This ordering of the magnetic field is also tied with the establishment of a circumbinary disc. We conclude the increased outflow velocity of T1 is due to this building-up of a circumbinary disc which then leads to the restructuring of the magnetic field to allow for efficient launching of outflows. In T2 the outflow velocities grow slowly before reaching a steady state at $\sim$3000$\yr$. From \Cref{fig:top_down_slices}, we see that the building of a circumbinary disc begins at earlier times for T2, with an extensive disc already visible at $\sim$3000$\yr$.

\begin{figure*}
	%\centerline{\includegraphics[width=1.0\linewidth]{Images/binary_system_time_evolution.pdf}}
	\centerline{\includegraphics[width=1.0\linewidth]{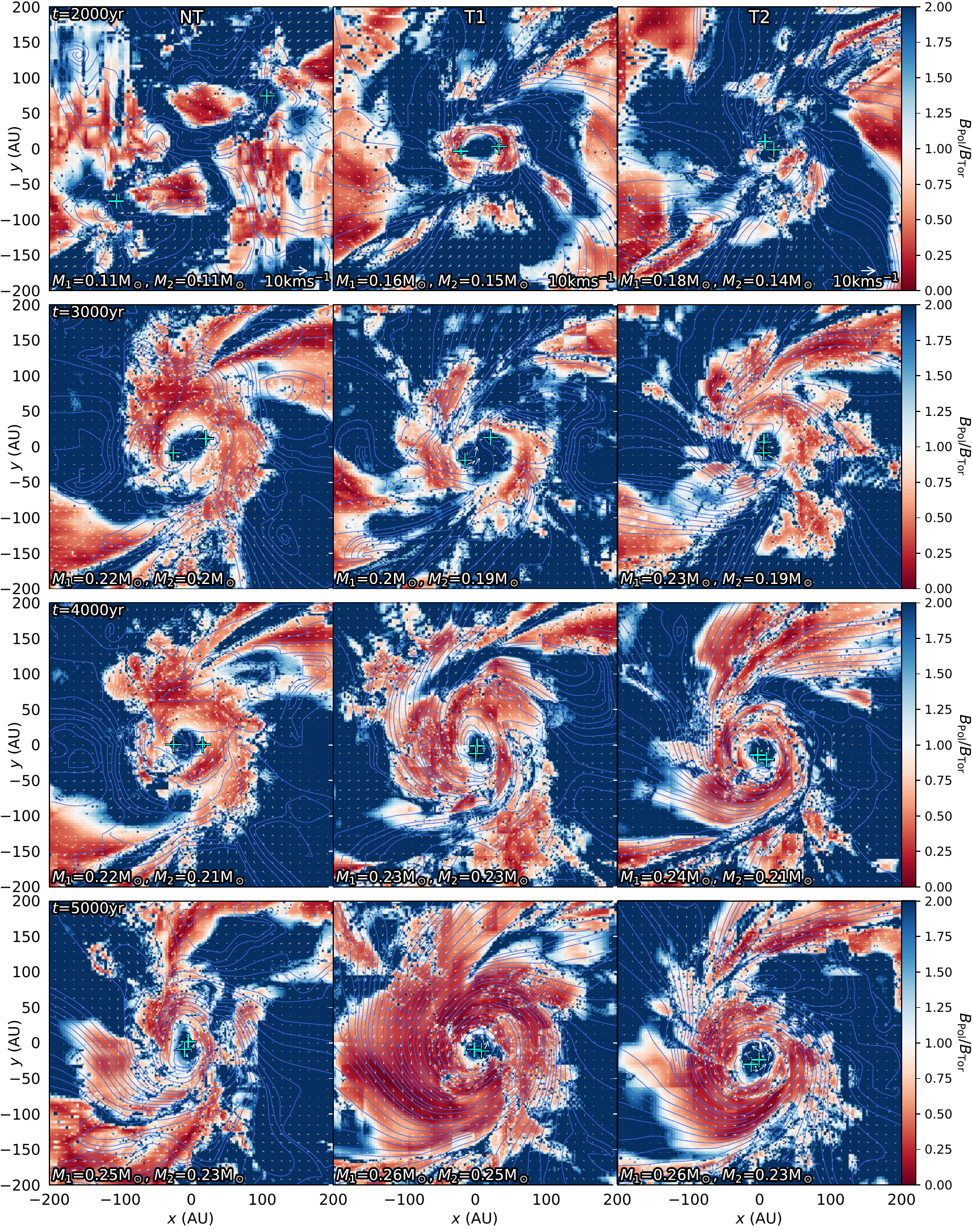}}
	%\caption{Shows the evolution of the binary separation (\emph{Top}), the total accretion rate (\emph{Middle}) and accreted mass (\emph{Bottom}) since protostar formation of the first sink particle for the NT, T1 and T2 cases in black, blue and red respective. In the plot of the accreted mass, the transparent thin lines trace the mass of the individual components and the thick opaque lines traces the total accreted mass.}
	\caption{Same as \Cref{fig:top_down_slices}, but shows volume-weighted slices of the ratio of the poloidal component ($B_\mathrm{pol}$) to the toroidal component ($B_\mathrm{tor}$) of the magnetic field.}
	\label{fig:pol_to_tor_ratio_xy}
\end{figure*}

\begin{figure*}
	\centering
	\centerline{\includegraphics[width=1.0\linewidth]{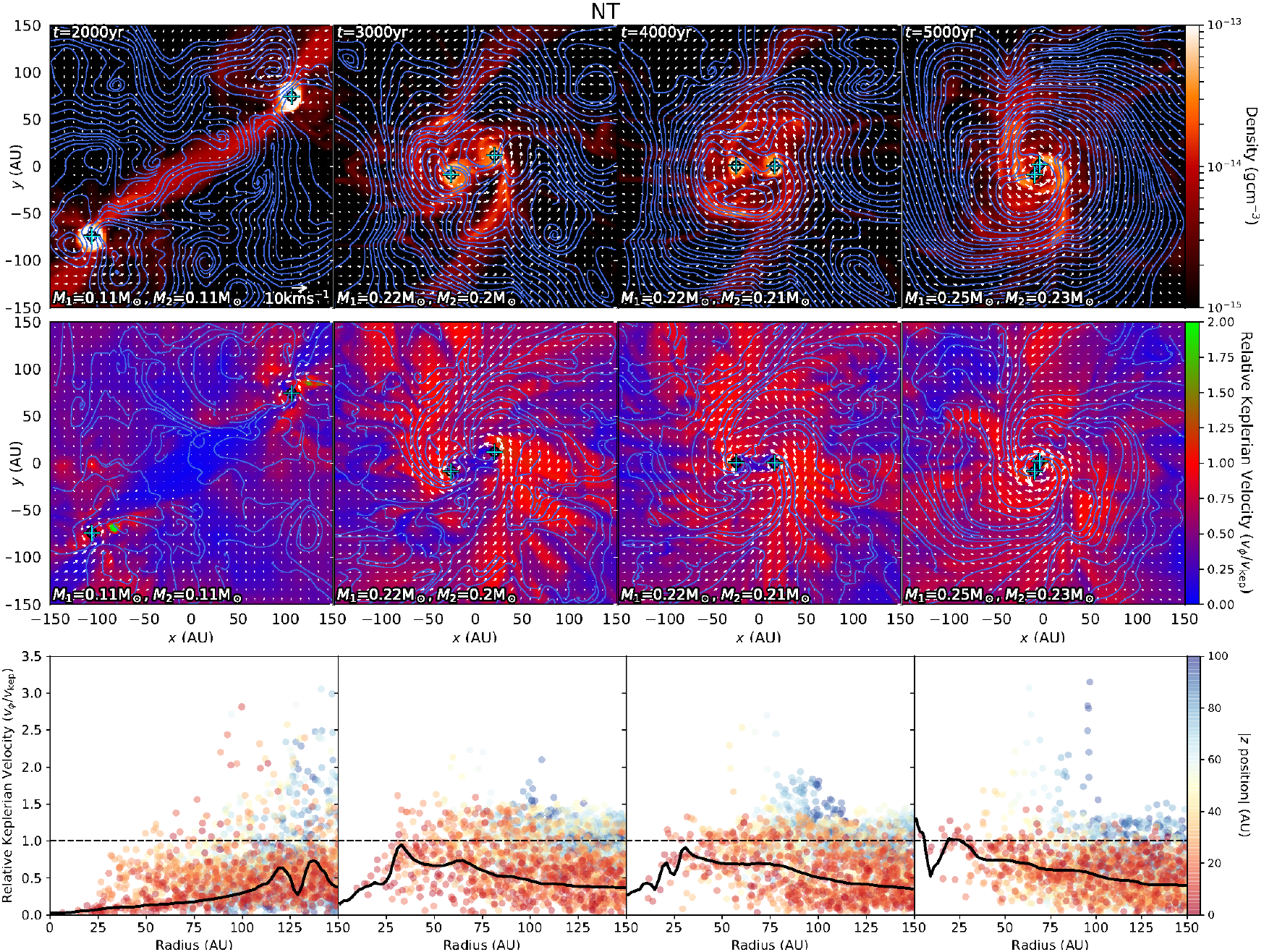}}
	%\resizebox{1.05\linewidth}{!}{\begin{tabular}{c}
	%		\hspace*{-1.1cm}\includegraphics[width=1.01\textwidth, trim={4mm, 4mm, 2mm, 2.5mm}, clip]{Images/Top_down_plots/Relative_Keplerian_Velocity/Mach_0.0/slices_xy_plane_unweighted_divided_by_thickness.pdf} \\
	%		\hspace*{-1.35cm}\includegraphics[width=1.003\textwidth, trim={2.5mm, 3.5mm, 3mm, 2mm}, clip]{Images/Top_down_plots/Relative_Keplerian_Velocity/Mach_0.0/profiles_no_lim.pdf} \\
	%\end{tabular}}
	\caption{Top and middle: $200\au$ thick volume-weighted projection plots of density and density-weighted projection plots of relative Keplerian velocity (described by Equation \ref{eqn:relativeKeplerianvelocity}) for NT. The thin lines show the magnetic field, and the arrows indicate the velocity field. The centre of the cross marks the position of the sink particle, and the black circle marks the accretion radius of the particle. Bottom: The solid black line is the density-weighted radial profile of the relative Keplerian velocity for a cylindrical volume of radius $150\au$ and thickness $200\au$ centred on the centre of mass. The points are randomly selected cells in the volume at various distances from the $z=0$ plane. Each column progresses at steps of $1000\yr$. The black dashed line highlights where material is Keplerian.}
	\label{fig:slices_mach_0.0}
\end{figure*}

\begin{figure*}
	\centering
	\centerline{\includegraphics[width=1.0\linewidth]{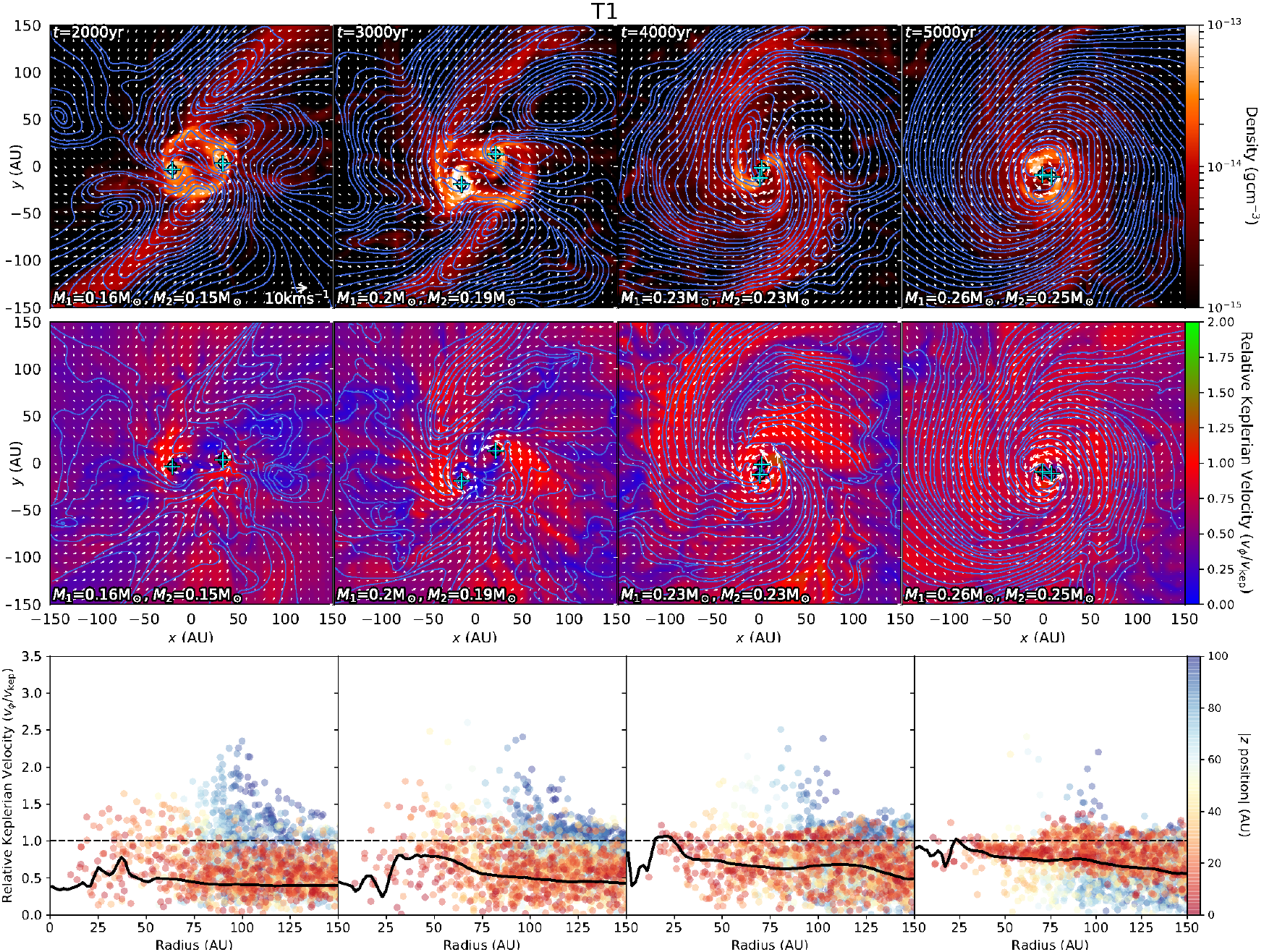}}
	%\resizebox{1.05\linewidth}{!}{\begin{tabular}{c}
	%		\hspace*{-1.1cm}\includegraphics[width=1.01\textwidth, trim={4mm, 4mm, 2mm, 2.5mm}, clip]{Images/Top_down_plots/Relative_Keplerian_Velocity/Mach_0.1/slices_xy_plane_unweighted_divided_by_thickness.pdf} \\
	%		\hspace*{-1.35cm}\includegraphics[width=1.003\textwidth, trim={2.5mm, 3.5mm, 3mm, 2mm}, clip]{Images/Top_down_plots/Relative_Keplerian_Velocity/Mach_0.1/profiles_no_lim.pdf} \\
	%\end{tabular}}
	\caption{Same as \Cref{fig:slices_mach_0.0}, but for T1.}
	\label{fig:slices_mach_0.1}
\end{figure*}

\begin{figure*}
	\centering
	\centerline{\includegraphics[width=1.0\linewidth]{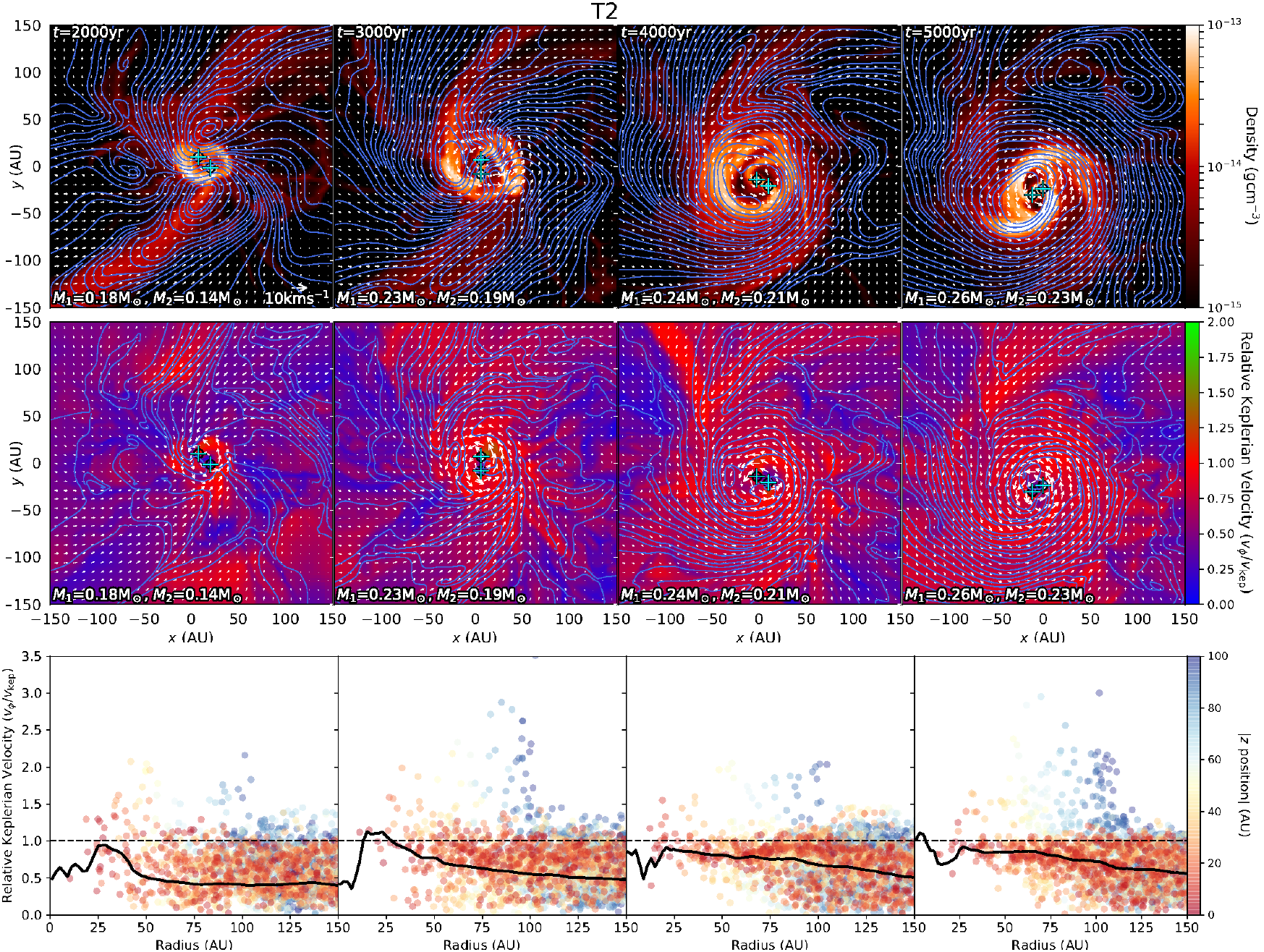}}
	%\resizebox{1.05\linewidth}{!}{\begin{tabular}{c}
	%		\hspace*{-1.1cm}\includegraphics[width=1.01\textwidth, trim={4mm, 4mm, 2mm, 2.5mm}, clip]{Images/Top_down_plots/Relative_Keplerian_Velocity/Mach_0.2/slices_xy_plane_unweighted_divided_by_thickness.pdf} \\
	%		\hspace*{-1.35cm}\includegraphics[width=1.003\textwidth, trim={2.5mm, 3.5mm, 3mm, 2mm}, clip]{Images/Top_down_plots/Relative_Keplerian_Velocity/Mach_0.2/profiles_no_lim.pdf} \\
	%\end{tabular}}
	\caption{Same as \Cref{fig:slices_mach_0.0}, but for T2.}
	\label{fig:slices_mach_0.2}
\end{figure*}

\subsection{Disc structure}
\label{ssec:discstructure}

We see from \Cref{fig:top_down_slices} that all the cases have the sink particles form and fall in along dense streams. As the sink particles fall towards each other we see dense circumstellar discs near the stars. As the binary systems in all the cases pass periastron the circumstellar discs are disrupted. As the circumstellar discs are disrupted, material is thrown outwards. Over the course of our simulations the turbulent cases appear to establish a large circumbinary disc. We see from \Cref{fig:top_down_slices} at later times ($\geq3000\yr$) that the turbulent simulations show the strong coiling of the magnetic field around the binary system as well as a dense circumbinary disc. The circumbinary disc is denser in T2 than in T1. The non-turbulent case does not appear to establish a circumbinary disc over the course of our simulation however, it might do so at later stages. As discussed in the previous section, the turbulence makes the disc more coherent. The formation of the disc restructures the magnetic field which then leads to the magnetic field also becoming more coherent. In \Cref{fig:pol_to_tor_ratio_xy} we show the ratio of the poloidal component to the toroidal component of the magnetic field for times $2000, 3000, 4000$ and $5000\,\mathrm{yr}$. As the simulations progress we see that the toroidal component dominates within the disc. The poloidal component seen at the midplane in \Cref{fig:pol_to_tor_ratio} is a result of the toroidal field changing direction. The size of the cavity that is predominately the poloidal component is approximately twice that of the binary star separation. The poloidal component is necessary for efficient launching of outflows and a strong toroidal component is needed for the collimation the launched outflows via the magneto-centrifugal mechanism \citep{blandford_hydromagnetic_1982}.

We determine how the relative Keplerian velocity of these circumbinary discs evolve to see if near-rotationally supported discs are created and how turbulence may influence this evolution. Rotationally supported gas discs can rotate at slightly sub-Keplerian velocities as the gas pressure contributes to supporting the disc against rapid infall \citep{adachi_gas_1976, weidenschilling_distribution_1977}. Studying the velocity structure throughout the discs provides details on mass transport around the forming binary star systems.

In Figures \ref{fig:slices_mach_0.0}, \ref{fig:slices_mach_0.1} and \ref{fig:slices_mach_0.2} we present $200\au$ thick volume-weighted projections along the $z=0$ plane of the gas discs of the density and density-weighted for the relative Keplerian velocity for our three simulations at $1000\yr$ intervals. We also show density-weighted radial profiles of the relative Keplerian velocity. The relative Keplerian velocity ($v_\mathrm{rel}$) is:

\begin{equation}
v_\mathrm{rel} = \frac{v_\phi}{v_\mathrm{kep}},
\label{eqn:relativeKeplerianvelocity}
\end{equation}

\noindent where $v_\phi$ is the tangential velocity to the radial direction and $v_\mathrm{kep}$ is the Keplerian velocity, given by: 

\begin{equation}
v_\mathrm{kep} = \sqrt{-\Phi_{\mathrm{tot}}}
\label{eqn:Keplerianvelocity}
\end{equation}

\noindent Where $\Phi_{\mathrm{tot}}$ is the total gravitational potential, i.e. the potential from the gas and the sink particles.

The density-weighted profiles (solid black line in bottom panel of Figures \ref{fig:slices_mach_0.0}, \ref{fig:slices_mach_0.1} and \ref{fig:slices_mach_0.2}) are calculated over a cylinder of radius $150\au$ and height $200\au$ centred on the centre of mass of the systems. The normal vector to this measuring cylinder is the angular momentum vector of the gas in the simulation domain. As most of the gas is within the discs this vector is perpendicular to the plane of the discs. $2\au$ cylindrical radius bins are used to produce these profiles. Along with the radial profile, randomly selected cells are plotted to show any variation in relative Keplerian velocity perpendicular to the disc.

In NT (\Cref{fig:slices_mach_0.0}), the evolution of the system is messy despite having no initial turbulence. At $2000\yr$ after formation of the first sink particle, the two stars have not passed the first periastron and they maintain their own circumstellar discs. At later times as the two stars interact, their circumstellar discs are disrupted and matter is ejected from the discs during the interaction. At time $5000\yr$ there is some circumbinary material that could hint at the formation of a circumbinary disc, however, only a very small one at this stage.

In the density projections, we clearly see that the magneticfield is disorganised at time $2000\yr$, but it begins to coil around the binary as the system evolves.

In the relative Keplerian velocity projections, we see most of the material in the vicinity is sub-Keplerian with a few patches of near Keplerian gas. The structure of the Keplerian material does not appear to follow the density structure closely.

The density-weighted profile plot from time $3000$ and $4000\yr$ show the velocity structure of the gas after the binary has completed a few orbits. At these times we see that the density-weighted profile is mostly sub-Keplerian, demonstrating that material is still in-falling. The material closest to the mid-plane (the red scatter points) is very sub-Keplerian, while material that is further away from the midplane (the blue scatter points) is more Keplerian to super-Keplerian. This is demonstrating the building up of the circumbinary environment but showing that gas falls into the system along the mid-plane and other gas is being carried away in the outflows. At later times it appears that material is being launched at radii up to $100\au$.

The density-weighted profile at time $5000\yr$ shows that there is a Keplerian disc out to $\sim$30$\au$. This rotationally supported disc may grow at later times.

In T1 (\Cref{fig:slices_mach_0.1}), the sink particles fall in towards each other faster than NT, and have already passed the first periastron by $2000\yr$. As the sink particles orbit each other material is thrown outwards and a circumbinary disc is beginning to build up at around $4000\yr$. By $5000\yr$ we see a circular circumbinary disc that has been built. The evolution of the magnetic field structure in this case is more pronounced compared to NT. By $4000\yr$ when we see the beginning of the circumbinary disc forming the magneticfield is orientated along the dense spiral arms that feed the disc. By $5000\yr$ the magnetic field has coiled around the binary system tightly. This restructuring of the magnetic field plays an important role in the magneto-centrifugal launching of outflows and in the increased outflow velocities discussed in \Cref{ssec:outflows} for the T1 case.

The relative Keplerian velocity slices for T1, like NT, show that most of the material in the vicinity of the stars is sub-Keplerian with a few patches of near Keplerian gas at times before the establishment of the circumbinary disc. However at later times we see at large radii the denser material
is very sub-Keplerian, and the gas in the disc is near Keplerian.

The density-weighted profiles for T1 begin sub-Keplerian and grow towards being near Keplerian. We see a gradient between sub-Keplerian mid-plane gas and Keplerian to super-Keplerian gas at larger $|z|$, similar to the NT case at these same times. At time $5000\yr$, the gas near the mid-plane is clustered near Keplerian our to $\sim$80$\au$. This is also the extent to which the circumbinary disc can be seen. Like NT, we see that material is being launched at radii up to $100\au$.

In T2 (\Cref{fig:slices_mach_0.2}), the circumbinary disc begins forming at earlier times compared to the other two cases. At time $3000\yr$, we see that a circumbinary disc has formed and continues to grow over the course of the simulation. We also see that the magnetic field coils about the binary system, but not as tightly as that seen in T1.

The density-weighted profiles for T2 follow the same trends as the other two cases. We see at later times that the mid-plane gas is sub-Keplerian to near Keplerian, and gas at large heights becomes more super-Keplerian. from the profile plot, the mostly Keplerian mid-plane extended to $80-100\au$. Also like the previous cases we see material being launched at radii up to $100\au$.

\subsubsection{Measuring disc sizes}
\label{sssec:disc_size}

\begin{figure*}
	\centering
	\centerline{\includegraphics[width=1.0\linewidth]{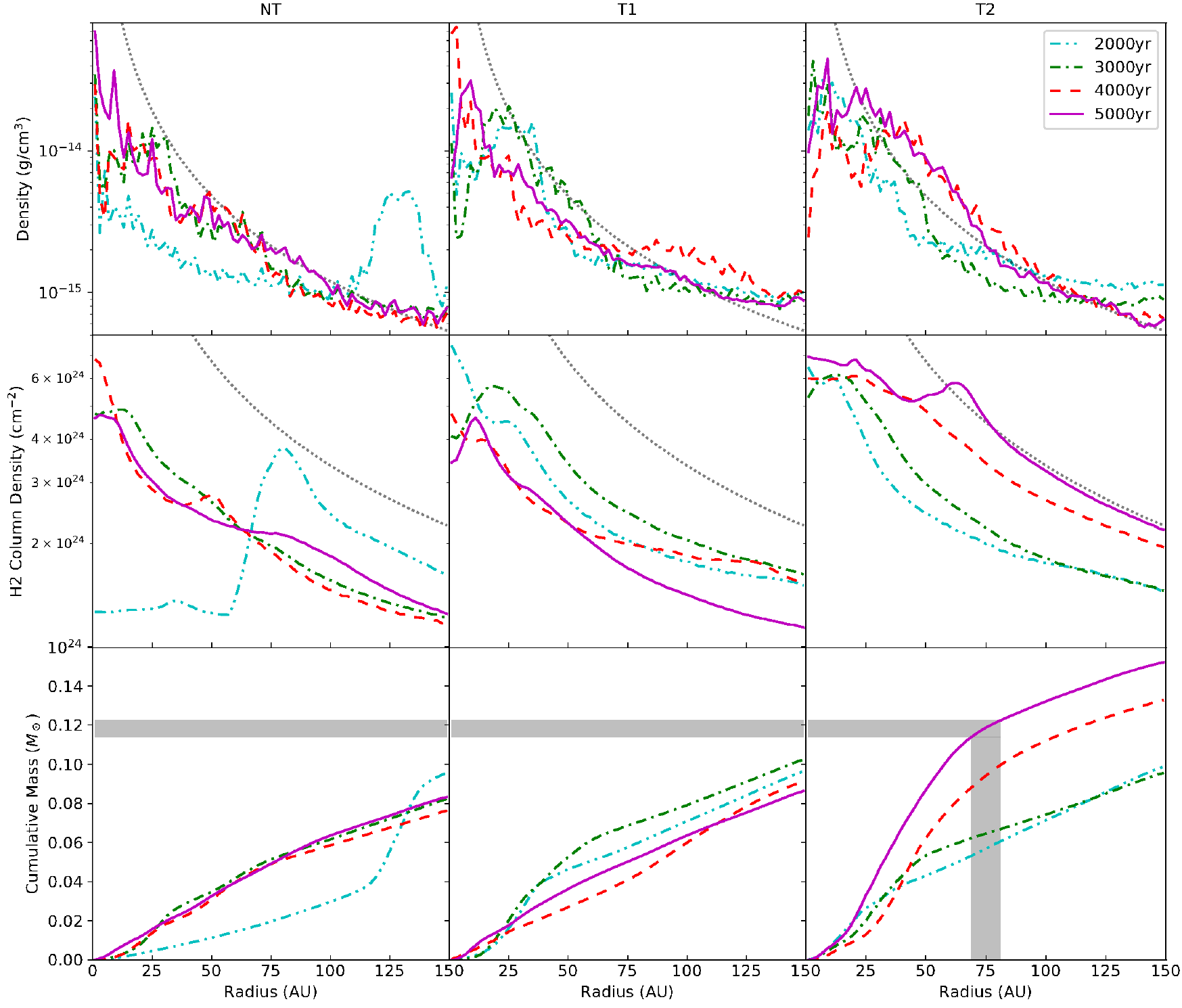}}
	\caption{Radial profiles of the volume-weighted mean gas density (top), integrated H2 column density (middle), as well as the cumulative gas mass (bottom) for the NT case (left), T1 case (middle) and T2 case (right) at times $2000\yr$ (dash-dash-dot cyan line), $3000\yr$ (dash-dot green line), $4000\yr$ (dashed red line) and $5000\yr$ (solid purple line) after sink particle formation. The dotted grey line in the top and middle panels highlight a $r^{-2}$ and $r^{-1}$ drop off in density and column density respectively. The grey region highlights the approximate radius and mass of the circumbinary disc in T2}
	\label{fig:density_profiles}
\end{figure*}

To compare the extent of circumbinary material we calculated radial profiles of the volume-weighted density, H2 column density and cumulative gas mass shown in \Cref{fig:density_profiles}. Density profiles are shown for $1000\yr$ intervals since $2000\yr$ after the formation of the first sink particle for our three cases. The measuring volume used to calculate the density profiles is the same used to calculate the relative Keplerian profiles shown in Figures \ref{fig:slices_mach_0.0}, \ref{fig:slices_mach_0.1} and \ref{fig:slices_mach_0.2}. %The top row of \Cref{fig:density_profiles} shows the volume-weighted radial density profiles. All cases show a background density of $<$10$^{-15}$~g~cm$^{-3}$. Determining the extent of a disc is ambiguous, therefore in this work we define the edge of the disc when the density profile drops to $\sim$5$\times 10^{-15}$~g~cm$^{-3}$ (indicated by the dashed grey line in \Cref{fig:density_profiles}). With this definition, at time $5000\yr$, the circumbinary discs have radii $\sim$30$\au$, $\sim$35$\au$, and $\sim$65$\au$ for NT, T1 and T2 respectively. The gas mass in the measuring cylinders out to the disc radii determined is $\sim$0.02$\msun$, $\sim$0.03$\msun$ and $\sim$0.11$\msun$ NT, T1 and T2 respectively.

From the overall shapes of the density profiles we see that the circumbinary material of NT does not show any significant growth over the course of the simulation after $3000\yr$. In T1 within $80\au$ the disc size oscillates, growing larger or smaller between the times plotted. Beyond $80\au$ we see that the density of the material oscillates out of phase to the growing and shrinking seen within $80\au$. This behaviour may hint at bulk movement of material at larger radii falling inwards. T2 is the only simulation that shows a clear inside-out growth of the circumbinary material, with the disc size steadily growing over the course of the simulation.

We also calculated H2 column density radial profiles to produce quantities measurable by observations. These are shown in the middle panels of \Cref{fig:density_profiles}. The H2 column density radial profiles are calculated from projection plots with the same thickness as the measuring volume used to calculate the relative Keplerian velocity and density profiles. The column density profiles also include the contribution of material above and below the bulk disc material (within the $\pm100\au$), which would also add to the line-of-sight density observed.

NT shows little growth after $3000\yr$, similar to what is shown by the density profiles. In T1, the column density increases between $2000$ and $3000\yr$, but then proceeds to decrease over the rest of the simulations. From \Cref{fig:outflow_quantities} it was between $3000$ and $4000\yr$ that the outflow velocities increase suddenly for T1. It is also between these times that we see the greatest decrease in column density. This is likely due to the increased outflow velocities suddenly clearing a cavity, reducing the column density. T2 is the only simulation that shows a clear inside-out growth in the column density of the circumbinary material, much like that seen in the density profiles. In the inner $\sim$75$\au$ we see perturbations (clumps, asymmetries) in the density / column density. These are the most prominent in the T2 case for the column density profile at $5000\yr$ where profile shows many features before dropping off beyond $75\au$. Features like these seen in observations can be indicative of turbulence producing rings, spiral arms and other disc characteristics \citep{hull_alma_2017, avenhaus_disks_2018, cox_almas_2018, esplin_wise_2018, sanna_discovery_2018, tychoniec_vla_2018}.

Determining the extent of the circumbinary disc is ambiguous, therefore in this work we look at the shape of the density and column density profiles. We approximate the disc radii to be where the profiles turn from showing features to dropping off smoothly with either a $r^{-2}$ profile for the density or $r^{-1}$ profile for the column density as seen in observations \citep{avenhaus_disks_2018}. Such power laws are indicated in \Cref{fig:density_profiles} by the dotted grey lines in the top and middle panels. From this we estimate the radii of the circumbinary discs at time $5000\yr$ to be $\lesssim20\au$, for NT and T1, and $\sim$70-80$\au$ for T2.

In order to estimate the mass of the circumbinary discs we have also calculated cumulative gas mass profiles, shown in the bottom panel of \Cref{fig:density_profiles}. From these profiles, at the radii estimated above, we find the circumbinary disc masses to be $\lesssim0.02\msun$ for NT and T1, and $\sim$0.12$\msun$ for T2. The cumulative mass profiles show the same evolution described by the other profiles above.

Figures \ref{fig:top_down_slices}, \ref{fig:slices_mach_0.1} and \ref{fig:slices_mach_0.2} also shows that at later times a minor inner hole in the circumbinary disc has been cleared by the binaries in the turbulent cases. The size of these inner holes is smaller or comparable in size to the semi-major axis of the binaries given by \Cref{fig:SystemEvolution}. This does not follow the prescription of \citet{artymowicz_dynamics_1994} who find the inner disc location to fall between $1.8a-2.6a$, where $a$ is the semi-major axis of the binary system. However these discs are still in the very early stages of formation.

\section{Limitations and caveats}
\label{sec:caveats}

\subsection{Numerical resolution}

The level of refinement used in our simulations does not resolve the regions closest to the actual protostar where high velocity outflows are launched. In our work, the resolution on the highest level of refinement ($L=12$) produces a cell size of $\Delta x\sim$1.95$\au$. \cite{federrath_modeling_2014} find that to have fully converged results for a simulation box size the same as that used in our work requires a level of refinement of $L=17$. This level refinement produces cell sizes of $\Delta x=0.06\au$. Running simulations with this level of refinement is very computationally intensive and impractical. This is the motivation for using $L=12$ for our work, as the primary goal is to compare relatively the influence of turbulence on binary star evolution. \citet{kuruwita_binary_2017} find using a level of refinement of $L=12$ is enough to resolve the larger scale outflows, while also allowing the simulations to evolve for a long time.

The evolution of accretion and the disc formation process is also sensitive to the initial and boundary conditions \citep{machida_conditions_2014,kuffmeier_zoom-simulations_2017, kuffmeier_episodic_2018}. \citet{machida_conditions_2014} find that the initial density distribution can affect the disc sized up to an order of magnitude. This was tested with a uniform density sphere and Bonner-Ebert profile. \citet{kuffmeier_zoom-simulations_2017, kuffmeier_episodic_2018} also find from zoom-in simulations that the density of the surrounding medium can affect the infall and accretion of the forming protostars. As our work is concerned with the relatively influence of turbulence on the evolution of discs in binaries, our results would not vary significantly as the initial and boundary conditions are consistent between all cases.

All of the binary star systems that formed in our simulations have separations of $\sim$10-20$\au$, which is resolved over $\sim$5-10 cells. Following the prescription of \cite{artymowicz_dynamics_1994}, if the sink particles host circumstellar discs, they would have a radius of $\sim$5$\au$. These discs would be resolved over approximately 5 cells. The contribution from these circumstellar discs is therefore, not well characterised at later times in our simulations. However, we are mostly concerned with the global evolution of these binary systems and the establishment of circumbinary discs. Using a higher level of refinement would allow us to resolve circumstellar discs around our sink particles, and understand more about their contribution to the global evolution of these binary systems.

\subsection{Radiation effects}

Our simulations do not explicitly calculate radiation transfer, however our EOS accounts for some radiative effects on the local cell scale (see \cref{ssec:flash}).

Radiation plays an important role in the initial collapse of molecular cores as it suppresses fragmentation \citep{krumholz_radiation-hydrodynamic_2007,offner_effects_2009,bate_stellar_2012,federrath_converging_2017}. Fragmentation of these cloud cores is believed one of the pathways in which binary stars and multiple star systems may form, therefore we should consider mechanisms that can influence fragmentation. Magnetic fields can also help suppress fragmentation by magnetically supporting the clouds \citep{price_effect_2008, burzle_protostellar_2011, peters_interplay_2011, federrath_star_2012, federrath_inefficient_2015, federrath_magnetic_2016}. However radiation-MHD simulations such as those conducted by \citet{myers_fragmentation_2013}, \citet{federrath_converging_2017}, and \citet{cunningham_effects_2018} find that radiation plays the most dominant role in fragmentation suppression in molecular cores.

Another pathway in which multiple star systems may form is via gravitational instability in discs. Simulations of binary star formation via this pathway find that even in relatively low-mass discs ($0.22\msun$) that once stellar cores form, the radiative feedback can suppress further disc fragmentation \citep{bate_stellar_2012}. Despite this radiative feedback effect, \citet{bate_stellar_2012} also conclude that gravity and gas dynamics are the main physical processes involved in determining the properties of multiple star systems.

From hydrodynamic simulations of star formation with radiative transfer in the flux-limited diffusion approximation \citet{offner_formation_2010} found that binary stars form predominantly via core fragmentation rather than disc instabilities. Therefore, of the two competing pathways in which binary systems may form, core fragmentation seems to be the more likely. 

%Once a protostar has formed, radiation continues to play a role in the evolution of the star and disc system. The effects of radiative feedback has mostly been investigated in the formation and evolution of massive star. Radiation hydrodynamic simulations of massive star formation find that radiative feedback dominates the star formation efficiency \citep{kuiper_protostellar_2016}. However with the inclusion of magnetic fields find that MHD disc winds are the dominant feedback mechanism \citealt{bate_collapse_2014, tanaka_impact_2017}. In these simulations of massive stars with radiation and magnetic field find that radiation continues to influence the star formation efficiency, despite not being the dominant mechanisms.

While most work on radiation feedback has focused on massive stars, some work has been done on low-mass stars. \citet{hansen_feedback_2012} ran 3D AMR radiation hydrodynamic simulations of low-mass star formation within a turbulent molecular core. This work also finds that stellar outflows dominate feedback reducing protostellar masses and accretion rates. A consequence of this is that the radiation pressure from the protostar is insignificant for low-mass stars, but radiation still suppresses fragmentation within the turbulent cloud by heating feedback.

Simulations of cluster formation incorporating both radiative feedback and ideal MHD have been carried out \citep{offner_effects_2009, price_inefficient_2009, myers_fragmentation_2013, myers_star_2014, krumholz_what_2016}. \citet{krumholz_what_2016} found that the formation of brown dwarfs of $\sim$0.01$\msun$ is greatly hindered by radiation. This is because thermal pressure in the vicinity of the protostar is increased due to accretion luminosity preventing further fragmentation \citep{federrath_converging_2017, guszejnov_isothermal_2018}. This may explain why low-mass binary stars are less frequent than high-mass binary stars.

Overall the effect that radiation may have on our simulations is that it would help suppress fragmentation. We believe the inclusion of radiation would not change the results of our simulations significantly as the stars formed are low-mass. However the influence of radiative feedback should be investigated to fully understand how the evolution of discs in binaries differs to disc evolution in single stars.

\subsection{Non-ideal MHD effects}

Non-ideal MHD effects are most important where partially ionised gas would exist such as in protoplanetary discs. The non-ideal effects are Ohmic resistivity, the Hall effect and ambipolar diffusion, and they are important at $\sim$1.5, $2-3$ and $\geq3$ scale heights from the disc-midplane respectively \citep{wardle_magnetic_2007, salmeron_magnetorotational_2008, konigl_effects_2011, tomida_radiation_2015, marchand_chemical_2016}. At scale heights greater than 3 the ideal MHD limit is a reasonable approximation. This is because the surface layers of discs are ionised by FUV stellar radiation \citep{perez-becker_surface_2011}. %However, \citet{wurster_effect_2018} argue that there is no realistic cosmic ray ionisation rate which would make star formation with non-ideal MHD converge to the ideal MHD limit. Ideal MHD also produces circumstellar magnetic fields a few orders of magnitude higher than observed for young stars; thus results from ideal MHD should be considered carefully \citep{wurster_origin_2018}.

The motivation to investigate non-ideal MHD effects on star formation is due to the magnetic breaking catastrophe, where magnetic fields are too efficient at breaking the rotation of the core in ideal MHD, suppressing the formation of circumstellar discs around young stars. Various works have found that non-ideal MHD effects can weaken the magnetic field strength and angular momentum transport helping to overcome magnetic breaking \citep{mellon_magnetic_2009, krasnopolsky_disk_2011, li_non-ideal_2011, dapp_bridging_2012, tsukamoto_effects_2015, wurster_can_2016,vaytet_protostellar_2018, zhao_decoupling_2018}. Because these non-ideal effects help to remove magnetic flux from magnetically sub-critical cores are able to collapse \citep{machida_different_2018}. \citet{wurster_collapse_2018} also find that the inclusion of all three non-ideal MHD effects along with radiation produced longer lived hydrostatic cores and second core phases. Applying this result to our simulations would delay the formation of sink particles. Despite delaying the collapse of molecular cores, non-ideal effects are also found to not significantly change the mass and radius of the first hydrostatic cores \citep{masson_ambipolar_2016}.

While most work on non-ideal MHD effects on star formation has mostly focused on single stars, there has been work on binary star formation \citep{duffin_early_2009}. \citet{machida_circumbinary_2009} found that Ohmic resistivity weakens the magnetic field strength near proto-binaries of separation $5-10\au$, but accumulated in the circumbinary disc where they are launched. \citet{wurster_impact_2017} found that the inclusion of Ohmic resistivity, the Hall effect and ambipolar diffusion in their simulations produced wider binaries with more massive discs. While they found that near periastron, non-ideal MHD effects were amplified, overall the initial conditions played the dominant role during binary star formation.

Our work does not consider non-ideal MHD effects because we are mainly concerned with the influence of turbulence on disc evolution in binary star systems. Given the results of previous studies on binary star formation the results of this study are not expected to change significantly with
the inclusion of non-ideal MHD effects. There is numerical diffusion in our simulations which weaken the magnetic field. In the turbulent cases we see circumbinary discs building up, which implies that turbulence plays a dominant role in angular momentum transport compared to the magnetic fields. Due to this, non-ideal MHD is not expected to significantly alter our main conclusions.

\section{Summary and conclusion}
\label{sec:conclusion}

We ran and analysed MHD simulations of binary star formation with varying levels of turbulence. We quantified the accretion, the outflow mass, momentum and angular momentum as well as the studied the morphology of outflows in three simulation cases: no turbulence (NT), Mach 0.1 turbulence (T1) and Mach 0.2 turbulence (T2). We also determined the evolution of the discs in these systems, in particular the building up of circumbinary discs. We find the following main results.

\emph{Orbital Evolution}. Turbulence produces asymmetric binaries. Stronger turbulence delays the absolute time taken for sink particles to form. This is likely due to the effective Jeans mass being greater with greater turbulence. Stronger turbulence also delays the formation of the secondary component of the binaries. The binary system formed in T2 circularises which is likely tied to the evolution of the associated circumbinary disc. While turbulence controls the star formation rate and efficiency on molecular cloud scales \citep{federrath_star_2012, krumholz_universal_2012, hennebelle_analytical_2013, padoan_infall-driven_2014}, we find here that it does not significantly influence the accretion rate of the stars inside the discs. We also see evidence for episodic accretion correlated with the periastron passage of the binary components (cf. \Cref{fig:top_down_slices} and 3).

\emph{Outflow Morphology}. All our simulations produce jets and outflows. The circumstellar discs around the protostars in NT and T1 launch individual jets before the stars complete a few orbits. The protostars in the most turbulent simulation fails to launch individual jets. This is likely due to the stronger turbulence disrupting the initial magnetic field structure, therefore producing a more unordered field, which does not allow for the magneto-centrifugal mechanism to work efficiently. At later times all systems produce a single outflow with the turbulent simulations producing outflows that appear bulkier and faster (cf. \Cref{fig:side_on_slices}).

More turbulent binary star formation can fail to launch individual outflows before coalescing. This may have interesting implications for observations as outflows are used as indicators of recent protostar birth. Because turbulent systems would not produce individual jets, the formation of binary systems in turbulent environments may be obscured to observations. However, this likely depends on the initial separation of the binary. Wider separation binaries formed in turbulent environments may produce individual outflows, however, this is outside the scope of this study.

\emph{Outflow Efficiency}. The non-turbulent simulation produces the most massive outflows, while the most turbulent simulation has the least massive outflows. However concerning linear and angular momentum transport, T1 produced outflow efficiencies comparable or more efficient than the non-turbulent case, while the most turbulent simulation was significantly less efficient at transporting momentum via outflows. Maximum outflow speeds remain relatively steady for the non-turbulence case, while the turbulent cases have maximum outflow speeds that grow by at least an order of magnitude over the course of the simulations (cf. \Cref{fig:outflow_quantities})

\emph{Disc Evolution}. All binaries host small circumstellar discs (approximately with a radius of $10\au$) which are destroyed as the system passes periastron and reform shortly thereafter. In the non-turbulent case the circumstellar discs do not survive long and the surrounding circumbinary material is diffuse and has little structure. In the turbulent cases the interaction of the binary stars redistributes angular momentum, such that a circumbinary disc can form faster than in the NT case. In the turbulent cases we also see that the magnetic fields coil up in the circumbinary discs (cf. \Cref{fig:top_down_slices}). This is what then primarily drives the outflows at late times. The T2 simulation produces the largest circumbinary disc (cf. \Cref{fig:density_profiles}). The circumbinary discs produced are rotationally supported.

Overall we find that turbulence helps to build up larger circumbinary discs, from which planets may form more easily than without the presence of turbulence.

\section*{Acknowledgements}

We thank the anonymous referee for insightful comments and improving this paper. R.K.~would like to thank the Australian Government and the financial support provided by the Research Training Program Domestic Scholarship. C.~F.~acknowledges funding provided by the Australian Research Council (Discovery Projects DP150104329 and DP170100603, and Future Fellowship FT180100495), and the Australia-Germany Joint Research Cooperation Scheme (UA-DAAD). The simulations presented in this work used high performance computing resources provided by the Leibniz Rechenzentrum and the Gauss Centre for Supercomputing (grants~pr32lo, pr48pi and GCS Large-scale project~10391), the Partnership for Advanced Computing in Europe (PRACE grant pr89mu), the Australian National Computational Infrastructure (grant~ek9), and the Pawsey Supercomputing Centre with funding from the Australian Government and the Government of Western Australia, in the framework of the National Computational Merit Allocation Scheme and the ANU Allocation Scheme. The simulation software \texttt{FLASH} was in part developed by the DOE-supported Flash Center for Computational Science at the University of Chicago. yt \citep{turk_yt:_2011} was used to help visualise and analyse these simulations.

\bibliographystyle{mn2e}
\bibliography{Bibliography}

\end{document}